\documentclass[journal,twocolumn]{IEEEtran}
\usepackage{graphicx}
\usepackage{amsmath,dsfont,bbm,epsfig,amssymb,amsfonts,  amstext, verbatim,amsopn,cite,subfigure,multirow,multicol,lipsum}
\usepackage{balance}
\usepackage{url}
\usepackage{amsfonts}
\usepackage{epsfig}
\usepackage{slashbox}
\usepackage{setspace}
\usepackage{stmaryrd}
\usepackage{psfrag}	
\usepackage{multirow}
\usepackage{float}
\usepackage{slashbox}
\usepackage[process=auto]{pstool}
\usepackage{etoolbox}
\usepackage{algorithm}
\usepackage{algorithmic}

\usepackage{pifont}
\allowdisplaybreaks

\newcommand*{\Scale}[2][4]{\scalebox{#1}{$#2$}}%

\newtheorem{lem}{Lemma}
\newtheorem{remk}{Remark}

\newtoggle{Extended}

\toggletrue{Extended}

\iftoggle{Extended}{%

}{%
  
}

\EndPreamble
\begin{document}

\title{Channel Estimation Techniques for Diffusion-Based Molecular Communications \vspace{-0.1cm}}

\author{Vahid Jamali\dag, Arman Ahmadzadeh\dag, Christophe Jardin\ddag, Heinrich Sticht\ddag, and Robert Schober\dag\\
\IEEEauthorblockA{\dag Institute for Digital Communications, \hspace{0.2cm}  \ddag Institute for Biochemistry \\
Friedrich-Alexander University (FAU), Erlangen, Germany \vspace{-0.3cm}
\thanks{This paper has been submitted for presentation at IEEE International Conference on Communications (ICC) 2016.}  
}
}

\maketitle

\begin{abstract}

In molecular communication (MC) systems, the \textit{expected} number of molecules observed at the receiver over time after the instantaneous release of molecules by the transmitter is referred to as the channel impulse response (CIR). Knowledge of the CIR is needed for the design of detection and equalization schemes. In this paper, we  present a training-based CIR estimation framework for MC systems which aims at estimating the CIR based on the \textit{observed} number of molecules at the receiver due to emission of a \textit{sequence} of known numbers of molecules by the transmitter.  In particular, we derive maximum likelihood (ML) and least sum of square errors (LSSE)  estimators. We also study the Cramer Rao (CR) lower bound and training sequence design for the considered system.  Simulation results confirm the analysis and compare the performance of the proposed estimation techniques with the CR lower bound. 
\end{abstract}

\section{Introduction}

Recent advances in biology, nanotechnology, and medicine have enabled the possibility of communication in nano/micrometer scale environments \cite{Survey_Mol_Net}.  Thereby,   employing molecules as information carriers, molecular communication (MC) has quickly emerged as a bio-inspired approach for man-made communication systems in such  environments. In fact, calcium signaling among neighboring cells, the use of neurotransmitters for communication across the synaptic cleft of neurons, and the exchange of autoinducers as signaling molecules in bacteria for quorum sensing are among the many examples of MC in nature \cite{Survey_Mol_Net}. 

\subsection{Motivation} 

 The design of any communication system crucially depends on the characteristics of the channel under consideration. In MC systems, the impact of the channel on
the number of observed molecules can be captured by the channel impulse response (CIR) which is defined as the \textit{expected} number of  molecules counted at the receiver at time $t$ after the instantaneous release of a known number of molecules by the transmitter at time $t=0$. The CIR, denoted by $\bar{c}(t)$, can be used as the basis for the design of   equalization and detection schemes for MC systems \cite{Hamid_Lett,Adam_OptReciever,ConsCIR}. For diffusion-based MC, the  released molecules move randomly according to Brownian motion  which is caused by thermal vibration and collisions with other  molecules in the fluid environment. Thereby, the average concentration of the molecules at a given coordinate $\mathbf{a}=[a_x, a_y, a_z]$ and at time $t$ after release by the transmitter, denoted by $\bar{\mathcal{C}}(\mathbf{a},t)$, is governed by Fick's second law of diffusion \cite{Adam_OptReciever}.  Finding $\bar{\mathcal{C}}(\mathbf{a},t)$ analytically involves solving partial differential equations and depends on initial and boundary conditions. Therefore, one possible approach for determining the CIR, which is widely employed in the literature \cite{ConsCIR}, is to first derive a sufficiently accurate analytical expression for $\bar{\mathcal{C}}(\mathbf{a},t)$ for the considered MC channel from Fick's second law, and to subsequently integrate it over the receiver volume, $V^{\mathtt{rec}}$, i.e., 
\begin{IEEEeqnarray}{lll} \label{Eq:Cons_CIR}
  \bar{c}(t) = \iiint_{\mathbf{a}\in V^{\mathtt{rec}}} \bar{\mathcal{C}}(\mathbf{a},t) \mathrm{d}a_x\mathrm{d}a_y\mathrm{d}a_z.
\end{IEEEeqnarray}
However, this approach may not be applicable in many practical scenarios as discussed in the following.

\begin{itemize}
\item The CIR can be obtained based on (\ref{Eq:Cons_CIR}) only for the special case of a fully \textit{transparent} receiver where it is assumed that the molecules move through the receiver as if it was not present in the environment. The assumption of a fully transparent receiver  is a valid approximation only for some particular scenarios where the interaction of the receiver with the molecules can be neglected. However, for general receivers, the relationship between the concentration $\bar{\mathcal{C}}(\mathbf{a},t)$  and the number of observed  molecules $\bar{c}(t)$ may not be as straightforward. 
\item Solving the differential equation associated with Fick's second law is possible only for simple and idealistic environments.  For example, assuming a \textit{point} source located at the origin of an \textit{unbounded} environment and \textit{impulsive} molecule release, $\bar{\mathcal{C}}(\mathbf{a},t)$ is obtained as \cite{ConsCIR}
\begin{IEEEeqnarray}{lll} \label{Eq:Consentration}
  \bar{\mathcal{C}}(\mathbf{a},t) = \frac{N^{\mathtt{TX}}}{\left(4\pi D t\right)^{3/2}} \exp\left(-\frac{|\mathbf{a}|^2}{4Dt}\right)\quad \left[\frac{\text{\small molecules}}{\text{\small m}^3}\right], \,\,\,
\end{IEEEeqnarray}
where $N^{\mathtt{TX}}$ is the number of molecules released  by the transmitter at $t=0$ and $D$ is the diffusion coefficient of the signaling molecule.  However, $\bar{\mathcal{C}}(\mathbf{a},t)$ cannot be obtained in closed form for most practical MC environments which may involve difficult boundary conditions, non-instantaneous molecule release, flow, etc.  Additionally, as has been shown in  \cite{Wil_Nature},  the classical Fick's diffusion equation  might even not be  applicable in  complex MC environments as  physicochemical interactions with other objects in the channel, such as other molecules, cells, and microvessels, are not accounted for.

\item Even if an expression for $\bar{\mathcal{C}}(\mathbf{a},t)$ can be obtained for a particular MC system, e.g. (\ref{Eq:Consentration}), it will be a function of several channel parameters such as the distance between the transmitter and the receiver and the diffusion coefficient. However, in practice, these parameters may  not be known a priori and also have to be estimated \cite{MC_Distance,AdamDistanceEstimation}. This complicates finding the CIR based on $\bar{\mathcal{C}}(\mathbf{a},t)$. 

\end{itemize}

Fortunately, for most communication problems, including equalization and detection, only the \textit{expected} number of molecules that the receiver observed at the sampling times is needed \cite{Adam_OptReciever,ConsCIR}. Therefore,  knowledge of how the average concentration is related to the channel parameters is not required, and hence, the difficulties associated with deriving $\bar{\mathcal{C}}(\mathbf{a},t)$ can be avoided by directly estimating the CIR. Analytical expressions for the CIR for specific assumptions for the transmitter, channel, and receiver are available in the literature. For example, the CIR for an unbounded environment and a  fully absorbing receiver  is given in \cite{Chae_Absorbing}. However, for general channel environments and  receivers, a simple closed-form expression for the expected number of observed molecules $\bar{c}(t)$ may not exist. Even if such an expression can be derived, it is only valid for a particular MC environment and is still a function of several unknown parameters.
 Motivated by the above discussion, our goal in this paper is to develop a general CIR estimation framework for  MC systems which is not limited to a particular MC channel model or a specific receiver type  and does not  require knowledge of the  channel parameters.

\subsection{Related Work} 

In most existing works on MC, the CIR is assumed to be perfectly known for  receiver design \cite{Hamid_Lett,Adam_OptReciever,ConsCIR,Arman_AF}. In the following, we  review the relevant MC literature that focused on channel characterization. Estimation of the distance between a transmitter and a receiver  was studied in \cite{MC_Distance,AdamDistanceEstimation} for diffusive MC. In \cite{Akyildiz_MC_E2E}, an end-to-end mathematical  model, including  transmitter,  channel, and  receiver, was presented, and in \cite{MC_Stoch_Model}, a stochastic channel model was proposed for  flow-based and diffusion-based MC. For active transport MC, a Markov chain channel model was derived in \cite{Farsad_Markov}. Additionally, a unifying model including the effects of external noise sources and inter-symbol interference (ISI) was proposed for diffusive MC  in \cite{Adam_Universal_Noise}.  In  \cite{Akyildiz_MC_Memory}, the authors analyzed a microfluidic MC channel, propagation noise, and channel memory. However, the focus  of \cite{MC_Distance,AdamDistanceEstimation,Akyildiz_MC_E2E,MC_Stoch_Model,Farsad_Markov,Adam_Universal_Noise,Akyildiz_MC_Memory} is either channel modeling or the estimation of channel parameters, i.e., the obtained results are not directly applicable to CIR acquisition. 

In contrast to MC, for conventional wireless communication, there is a rich literature on channel estimation, mainly for linear channel models and impairment by additive white Gaussian noise (AWGN), see \cite{LSSE,Channel_Estimation_Arsalan}, and the references therein.  Channel estimation was also studied for non-linear and/or non-AWGN channels especially  in  optical communication. For example, for a photon-counting receiver, a linear time-invariant channel model with Poisson noise was considered in \cite{Channel_Estimation_Optic_ISI} and a non-linear channel model  with Poisson noise was  investigated in \cite{Channel_Estimation_Optic_Squ}. However, the  MC channel model considered in this paper is neither linear  nor impaired by AWGN and is also different from that in \cite{Channel_Estimation_Optic_Squ}. Therefore, the results known from conventional wireless communication are not directly applicable to MC.

\subsection{Contributions} 
 
In contrast to \cite{MC_Distance,Arman_AF,Chae_Absorbing,AdamDistanceEstimation,Akyildiz_MC_E2E,MC_Stoch_Model,Farsad_Markov,Adam_Universal_Noise,Akyildiz_MC_Memory}, in this paper,  we directly estimate the CIR based on the channel output, i.e., the number of molecules observed at the receiver.  To the best of the authors' knowledge, this problem has not been studied  in the MC literature, yet.  In particular, we present a training-based CIR estimation framework which aims at estimating the CIR based on the detected number of molecules at the receiver due to the emission of a  sequence of known numbers of molecules by the transmitter. To this end, we first derive the optimal maximum likelihood (ML) CIR  estimator. Subsequently, we obtain the suboptimal least sum of square errors (LSSE) CIR  estimator which entails a lower computational complexity than the ML estimator. Additionally, we derive the Cramer Rao (CR) bound which constitutes a lower bound on the estimation error variance of any unbiased estimator. We also study training sequence design for the considered MC system. Simulation results  confirm the analysis and evaluate the performance of the proposed estimation techniques with respect to  the CR lower bound. 
  
\textit{Notations:} We use the following notations throughout this paper: $\mathbbmss{E}_{x}\{\cdot\}$ denotes expectation with respect to random variable (RV) $x$ and $[x]^+=\max\{0,x\}$.  Bold capital and small letters are used to denote matrices and vectors, respectively. $\mathbf{1}$ and $\mathbf{0}$ are vectors whose elements are all ones and zeros, respectively, $\mathbf{A}^T$ denotes the transpose of $\mathbf{A}$,  $\|\mathbf{a}\|$ represents the norm of  vector $\mathbf{a}$, $[\mathbf{A}]_{mn}$ denotes the element in the $m$-th row and $n$-th column of matrix $\mathbf{A}$, $\mathrm{tr}\{\mathbf{A}\}$ is the trace of matrix $\mathbf{A}$, $\mathrm{diag}\{\mathbf{a}\}$ denotes a diagonal matrix with the elements of vector $\mathbf{a}$ on its main diagonal, $\mathrm{vdiag}\{\mathbf{A}\}$ is a vector which contains the diagonal entries
of  matrix $\mathbf{A}$, $\mathrm{eig}\{\mathbf{A}\}$ is the set of eigen-values of matrix $\mathbf{A}$,  $\mathbf{A}\succeq 0$  denotes a positive semidefinite matrix $\mathbf{A}$, and $\mathbf{a}\geq \mathbf{0}$ means that all the elements of vector $\mathbf{a}$ are non-negative. Additionally, $\mathrm{Poiss}(\lambda)$  denotes a Poisson RV with mean  $\lambda$, and $\mathrm{Bin}(n,p)$ denotes a binomial RV for $n$  trials and success probability $p$.

\section{Problem Formulation}

In this section, we first present the considered MC channel model, and subsequently, formulate the CIR estimation problem.  

\subsection{System Model}

We consider an MC system  consisting of a transmitter, a channel, and a receiver. At the beginning of each symbol interval, the transmitter releases either $N^{\mathtt{TX}}$ or zero molecules, i.e., ON-OFF keying is performed. In this paper, we assume that the transmitter emits only one type of molecule. The released molecules propagate through the medium between the transmitter and the receiver. We assume that the movements of individual molecules are independent from each other. The receiver counts the number of observed molecules in each symbol interval.  We note that this is a rather general model for the MC receiver which includes well-known receivers such as the transparent receiver \cite{ConsCIR} and the absorbing receiver \cite{Chae_Absorbing}.

 
Due to the memory of the MC channel, inter-symbol interference (ISI) occurs \cite{Adam_Universal_Noise,Akyildiz_MC_Memory}. In particular, ISI-free communication is only possible  if  the symbol intervals are chosen sufficiently large such that the CIR fully decays to zero within one symbol interval which severely limits the transmission rate and results in an inefficient MC system design. Therefore, taking into account the effect of ISI, we assume the following input-output relation for the MC system
\begin{IEEEeqnarray}{lll} \label{Eq:ChannelInOut}
  r[k]  = \sum_{l=1}^{L} c_l[k] + c_{\mathtt{n}}[k],
\end{IEEEeqnarray}
where $r[k]$ is the number of molecules detected at the receiver in symbol interval $k$, $L$ is the number of memory taps of the channel, and $c_l[k]$ is the number of  molecules observed at the receiver in symbol interval $k$ due to the release of $s[k-l+1]N^{\mathtt{Tx}}$ molecules by the transmitter in symbol interval $k-l+1$, where $s[k]\in\{0,1\}$ holds.
Thereby,  $c_l[k]$ can be well approximated by a Poisson RV with mean $\bar{c}_l s[k-l+1]$, i.e., $c_l[k]\sim\mathrm{Poiss}\left(\bar{c}_l s[k-l+1]\right)$, see \cite{Hamid_Lett,Adam_OptReciever}. Moreover, $c_{\mathtt{n}}[k]$ is the number of external noise molecules detected by the receiver in symbol interval $k$  but not released by the transmitter. Noise molecules may originate from  interfering sources which employ the same type of molecule as the considered MC system. Hence, $c_{\mathtt{n}}[k]$ can  also be modeled as a Poisson  RV, i.e., $c_{\mathtt{n}}[k]\sim\mathrm{Poiss}\left(\bar{c}_{\mathtt{n}}\right)$, where $\bar{c}_{\mathtt{n}}=\mathbbmss{E}\left\{c_{\mathtt{n}}[k]\right\}$. 

\begin{remk}
From a probabilistic point of view, we can assume that each molecule released by the transmitter in symbol interval $k-l+1$ is observed at the receiver in symbol interval $k$ with a certain probability, denoted by $p_l$. Thereby, the probability that $n$ molecules are observed at the receiver  in symbol interval $k$ due to the emission of $N^{\mathtt{Tx}}$ molecules in symbol interval $k-l+1$ follows a binomial distribution, i.e., $n\sim\mathrm{Bin}(N^{\mathtt{Tx}},p_l)$. Moreover, assuming $N^{\mathtt{Tx}}\to\infty$ while $N^{\mathtt{Tx}}p_l\triangleq \bar{c}_l$ is fixed, the  binomial distribution $\mathrm{Bin}(N^{\mathtt{Tx}},p_l)$ converges to the Poisson distribution $\mathrm{Poiss}(\bar{c}_l)$ \cite{BayesianBook}. This  is a  valid assumption in MC since the number of released molecules is often very large to ensure that a sufficient number of molecules reaches the receiver. The same reasoning applies to the noise molecules.
\end{remk}

 Unlike the conventional linear input-output  model for channels with memory in wireless communication systems \cite{LSSE,Channel_Estimation_Arsalan}, the channel model in (\ref{Eq:ChannelInOut}) is not linear since $s[k-l+1]$ does not affect the observation $r[k]$ directly but via Poisson RV $c_l[k]$. However, the \textit{expectation} of the received signal is linearly dependent on the transmitted signal, i.e.,
\begin{IEEEeqnarray}{lll} \label{Eq:AveInOut}
 \bar{r}[k] = \mathbbmss{E}\left\{r[k]\right\} = \sum_{l=1}^{L} \bar{c}_l s[k-l+1] + \bar{c}_{\mathtt{n}}.
\end{IEEEeqnarray}
We note that for a given $s[k]$, in general, the actual number of molecules observed at the receiver, $r[k]$, will differ from the expected number of observed molecules, $\bar{r}[k]$, due to the intrinsic noisiness of diffusion.

\subsection{CIR Estimation Problem}

  Let $\mathbf{s}=[s[1],s[2],\dots,s[K]]^T$ be a training sequence of length $K$. Here, we assume continuous transmission. Therefore, in order to ensure that the received signal is only affected by the training sequence $\mathbf{s}$ and not by the  transmissions in previous symbol intervals, we only employ $r[k],\,\,k\geq L$, for CIR estimation. Thereby, the $K-L+1$   samples  used for CIR estimation are given by
\begin{IEEEeqnarray}{lll} \label{Eq:SysRecMol}
 r[L]   =  \mathrm{Poiss}\left(\bar{c}_1  s[L]\right) + \mathrm{Poiss}\left(\bar{c}_2  s[L-1]\right) + \cdots   \nonumber \\
\qquad\qquad\qquad\qquad\qquad \,   + \mathrm{Poiss}\left(\bar{c}_L  s[1]\right) + \mathrm{Poiss}\left(\bar{c}_{\mathtt{n}}\right)   \quad\,\,\,\, \IEEEyesnumber\IEEEyessubnumber \\
  r[L+1]   =  \mathrm{Poiss}\left(\bar{c}_1  s[L+1]\right) + \mathrm{Poiss}\left(\bar{c}_2  s[L]\right) + \cdots   \nonumber \\
\qquad\qquad\qquad\qquad\qquad \,   + \mathrm{Poiss}\left(\bar{c}_L  s[2]\right) + \mathrm{Poiss}\left(\bar{c}_{\mathtt{n}}\right)   \IEEEyessubnumber \\
\qquad \Scale[0.95]{  \vdots } \qquad\qquad\qquad\qquad\qquad \Scale[0.95]{  \vdots }\nonumber \\
 r[K]   =  \mathrm{Poiss}\left(\bar{c}_1  s[K]\right) + \mathrm{Poiss}\left(\bar{c}_2  s[K-1]\right) + \cdots   \nonumber \\
 \qquad\qquad\qquad \,   + \mathrm{Poiss}\left(\bar{c}_L  s[K-L+1]\right) + \mathrm{Poiss}\left(\bar{c}_{\mathtt{n}}\right).   \qquad \IEEEyessubnumber 
\end{IEEEeqnarray}
For convenience of notation, we define $\mathbf{r}=[r[L],r[L+1],\dots,r[K]]^T$ and $\bar{\mathbf{c}}=[\bar{c}_1, \bar{c}_2,\dots,\bar{c}_L,\bar{c}_{\mathtt{n}}]^T$, and $f_{\mathbf{r}}(\mathbf{r}|\bar{\mathbf{c}},\mathbf{s})$ is the probability density distribution (PDF) of observation  $\mathbf{r}$  conditioned on a given channel $\bar{\mathbf{c}}$ and a given training sequence  $\mathbf{s}$. We assume that the CIR\footnote{With a slight abuse of notation, in the following, we refer to vector $\bar{\mathbf{c}}$ as the CIR although $\bar{\mathbf{c}}$ also contains the mean of the noise $\bar{c}_{\mathtt{n}}$.}, $\bar{\mathbf{c}}$, remains unchanged for a sufficiently large block of  symbol intervals during which CIR estimation and data transmission are performed. However, the CIR may change from one block to the next due to e.g. a change in the distance between transmitter and receiver.   To summarize,  in each block, the stochastic model in (\ref{Eq:ChannelInOut}) is characterized by $\bar{\mathbf{c}}$ and  our goal in this paper is to estimate $\bar{\mathbf{c}}$ based on the vector of random observations $\mathbf{r}$.


\section{CIR Estimation}

In this section, we derive the ML and LSSE  estimators as well as the CR lower bound for CIR estimation in MC.

\subsection{ML CIR Estimation}

The ML CIR estimator chooses the CIR which maximizes the likelihood of  observation vector $\mathbf{r}$ \cite{BayesianBook}. In particular,  the ML estimator is given by
\begin{IEEEeqnarray}{lll} \label{Eq:ML_Estimation}
  \hat{\bar{\mathbf{c}}}^{\mathtt{ML}} = \underset{\bar{\mathbf{c}}\geq \mathbf{0}}{\mathrm{argmax}} \,\,f_{\mathbf{r}}(\mathbf{r}|\bar{\mathbf{c}},\mathbf{s}).
\end{IEEEeqnarray}
We assume that the observations in different symbol intervals are independent, i.e., $r[k]$ is independent of $r[k']$ for $k\neq k'$. This assumption is valid in practice if the time interval between two consecutive samples is sufficiently large, see \cite{Adam_OptReciever} for a detailed discussion. Moreover, from (\ref{Eq:ChannelInOut}), we observe that $r[k]$ is a sum of  Poisson RVs. Hence, $r[k]$ is also a Poisson RV with its mean equal to the sum of the means of the summands, i.e., $r[k]\sim\mathrm{Poiss}(\bar{r}[k])$ with $\bar{r}[k] = \bar{c}_{\mathtt{n}} + \sum_{l=1}^{L} \bar{c}_l s[k-l+1] = \bar{\mathbf{c}}^T\mathbf{s}_k  $ and $\mathbf{s}_k=[s[k],s[k-1],\dots,s[k-L+1],1]^T$.  Therefore, $f_{\mathbf{r}}(\mathbf{r}|\bar{\mathbf{c}},\mathbf{s})$ is given by
\begin{IEEEeqnarray}{rll} \label{Eq:ML_PDF}
 f_{\mathbf{r}}(\mathbf{r}|\bar{\mathbf{c}},\mathbf{s})\,\, 
& = \prod_{k=L}^{K} \frac{\left(\bar{\mathbf{c}}^T\mathbf{s}_k  \right)^{r[k]} \exp\left(-\bar{\mathbf{c}}^T\mathbf{s}_k  \right)}{r[k]!}. 
\end{IEEEeqnarray}
 Maximizing $f_{\mathbf{r}}(\mathbf{r}|\bar{\mathbf{c}},\mathbf{s})$ is equivalent to maximizing $\mathrm{ln}(f_{\mathbf{r}}(\mathbf{r}|\bar{\mathbf{c}},\mathbf{s}))$ since $\mathrm{ln}(\cdot)$ is a monotonically increasing function. Hence, the ML estimate  can be rewritten as
\begin{IEEEeqnarray}{cll} \label{Eq:ML_Log}
  \hat{\bar{\mathbf{c}}}^{\mathtt{ML}} = \underset{\bar{\mathbf{c}}\geq \mathbf{0}}{\mathrm{argmax}} \,\,g(\bar{\mathbf{c}}) \quad \mathrm{where}   \\ 
 g(\bar{\mathbf{c}}) \triangleq \sum_{k=L}^{K} \Big[-\bar{\mathbf{c}}^T\mathbf{s}_k   + r[k]\mathrm{ln}\left(\bar{\mathbf{c}}^T\mathbf{s}_k  \right)\Big].\nonumber
\end{IEEEeqnarray}
To present the solution of the above optimization problem rigorously, we first define some auxiliary variables. Let $\mathcal{A}=\{\mathcal{A}_1,\mathcal{A}_2,\dots,\mathcal{A}_N\}$ denote a set which contains all possible $N=2^{L+1}-1$ subsets of set $\mathcal{F}=\{1,2,\cdots,L,\mathtt{n}\}$ except the empty set. Here, $\mathcal{A}_n,\,\,n=1,2,\dots,N$, denotes the $n$-th subset  of $\mathcal{A}$. Moreover, let $\bar{\mathbf{c}}^{\mathcal{A}_n}$ and $\mathbf{s}_k^{\mathcal{A}_n}$ denote reduced-dimension versions of  $\bar{\mathbf{c}}$ and $\mathbf{s}_k$, respectively, which contain only the elements of $\bar{\mathbf{c}}$ and $\mathbf{s}_k$ whose indices are in set $\mathcal{A}_n$, respectively.

\begin{lem}\label{Lem:ML}
The ML estimator of the CIR for the considered MC channel is given by Algorithm~1 where the following non-linear system of equations is solved\footnote{The system of nonlinear equations in (\ref{Eq:ML_Sol}) can be solved using standard mathematical software packages such as Mathematica.} for different $\mathcal{A}_n$
\begin{IEEEeqnarray}{lll} \label{Eq:ML_Sol}
\sum_{k=L}^{K} \left[\frac{r[k] }{(\bar{\mathbf{c}}^{\mathcal{A}_n})^T \mathbf{s}_k^{\mathcal{A}_n}} -1\right]  \mathbf{s}_k^{\mathcal{A}_n} = \mathbf{0}.
\end{IEEEeqnarray}
\end{lem}

\begin{IEEEproof}
The problem in (\ref{Eq:ML_Log}) is a convex optimization problem in variable $\bar{\mathbf{c}}$ because $g(\bar{\mathbf{c}})$ is a concave function in $\bar{\mathbf{c}}$ and the feasible set $\bar{\mathbf{c}}\geq \mathbf{0}$ is linear in $\bar{\mathbf{c}}$.  In particular, $\mathrm{ln}\left(\bar{\mathbf{c}}^T\mathbf{s}_k  \right)$ is concave  since $\bar{\mathbf{c}}^T\mathbf{s}_k$ is affine and the log-function is concave \cite[Chapter~3]{Boyd}. Therefore,  $g(\bar{\mathbf{c}})$ is a sum of weighted concave terms $r[k]\mathrm{ln}\left(\bar{\mathbf{c}}^T\mathbf{s}_k  \right)$ and affine terms $\bar{\mathbf{c}}^T\mathbf{s}_k$ which in turn yields a concave function \cite[Chapter~3]{Boyd}. For the constrained convex problem in (\ref{Eq:ML_Log}), the optimal solution falls into one of the following two categories:

\textit{Stationary Point:} In this case, the optimal solution is found by taking the derivative of  $g(\bar{\mathbf{c}})$ with respect to $\bar{\mathbf{c}}$ and setting $\bar{\mathbf{c}}^{\mathcal{F}}=\bar{\mathbf{c}}$ and $\mathbf{s}_k^{\mathcal{F}}=\mathbf{s}_k$ which leads to (\ref{Eq:ML_Sol}) for $\mathcal{A}_n=\mathcal{F}$. Note that this stationary point is the global optimal solution of the unconstrained version of the problem in (\ref{Eq:ML_Log}), i.e., when constraint $\bar{\mathbf{c}}\geq\mathbf{0}$ is dropped. Therefore, if $\bar{\mathbf{c}}^{\mathcal{F}}$ is in the feasible set, i.e., $\bar{\mathbf{c}}^{\mathcal{F}}\geq\mathbf{0}$ holds, it is also the optimal solution of the constrained problem in (\ref{Eq:ML_Log}) and hence, we obtain $\hat{\bar{\mathbf{c}}}^{\mathtt{ML}}=\bar{\mathbf{c}}^{\mathcal{F}}$.

\textit{Boundary Point:} In this case, for the optimal solution, some of the elements of $\bar{\mathbf{c}}$ are zero. Since it is not a priori known which elements are zero, we have to consider all possible cases. To do so, we use auxiliary variables $\bar{\mathbf{c}}^{\mathcal{A}_n}$ and $\mathbf{s}_k^{\mathcal{A}_n}$ where set $\mathcal{A}_n$ specifies the indices of the non-zero elements of $\bar{\mathbf{c}}$. For a given $\mathcal{A}_n$, we formulate a new problem by substituting $\bar{\mathbf{c}}^{\mathcal{A}_n}$ and $\mathbf{s}_k^{\mathcal{A}_n}$ for $\bar{\mathbf{c}}$ and $\mathbf{s}_k$ in (\ref{Eq:ML_Log}), respectively. The solution of the \textit{new problem} is now a stationary point not a boundary point since a boundary point implies that some of the elements of $\bar{\mathbf{c}}^{\mathcal{A}_n}$ are zero which yields a contradiction because we assumed that $\bar{\mathbf{c}}^{\mathcal{A}_n}$ includes the non-zero elements of $\bar{\mathbf{c}}$. The stationary point of the new problem can be found by taking the derivative of  $g(\bar{\mathbf{c}}^{\mathcal{A}_n})$ with respect to $\bar{\mathbf{c}}^{\mathcal{A}_n}$ which leads to (\ref{Eq:ML_Sol}). Here, if $\bar{\mathbf{c}}^{\mathcal{A}_n}\geq \mathbf{0}$ does not hold, we discard $\bar{\mathbf{c}}^{\mathcal{A}_n}$, otherwise, it is a candidate for the optimal solution. Therefore, we  construct the candidate ML CIR estimate, denoted by $\hat{\bar{\mathbf{c}}}^{\mathtt{CAN}}$, such that the elements of $\hat{\bar{\mathbf{c}}}^{\mathtt{CAN}}$ whose indices are in $\mathcal{A}_n$ are equal to the values of the corresponding elements in $\bar{\mathbf{c}}^{\mathcal{A}_n}$ and the remaining elements are equal to zero. The resulting $\hat{\bar{\mathbf{c}}}^{\mathtt{CAN}}$ is saved in the candidate set $\mathcal{C}$. Finally, the ML estimate, $\hat{\bar{\mathbf{c}}}^{\mathtt{ML}}$, is given by that $\hat{\bar{\mathbf{c}}}^{\mathtt{CAN}}$ in set $\mathcal{C}$ which maximizes $g(\bar{\mathbf{c}})$.

The above results are concisely summarized in Algorithm~1 which concludes the proof.
\end{IEEEproof}

\begin{algorithm}[t] 
{\fontsize{10}{10}\selectfont
\caption{ {\fontsize{10}{10}\selectfont
{\color[rgb]{0,0,1}ML}/{\color[rgb]{1,0,0}LSSE} CIR Estimate {\color[rgb]{0,0,1}$\hat{\bar{\mathbf{c}}}^{\mathtt{ML}}$}/{\color[rgb]{1,0,0}$\hat{\bar{\mathbf{c}}}^{\mathtt{LSSE}}$}
} }
\begin{algorithmic} 
\STATE \textbf{initialize} $\mathcal{A}_n=\mathcal{F}$ and solve {\color[rgb]{0,0,1}(\ref{Eq:ML_Sol})}/{\color[rgb]{1,0,0}(\ref{Eq:LSSE_Sol})}  to find $\bar{\mathbf{c}}^{\mathcal{F}}$ 
\IF{$\bar{\mathbf{c}}^{\mathcal{F}}\geq \mathbf{0}$}
\STATE  Set {\color[rgb]{0,0,1}$\hat{\bar{\mathbf{c}}}^{\mathtt{ML}} = \bar{\mathbf{c}}^{\mathcal{F}}$}/{\color[rgb]{1,0,0}$\hat{\bar{\mathbf{c}}}^{\mathtt{LSSE}} = \bar{\mathbf{c}}^{\mathcal{F}}$}
\ELSE
 \FOR{$\forall \mathcal{A}_n\neq \mathcal{F}$}
        \STATE Solve {\color[rgb]{0,0,1}(\ref{Eq:ML_Sol})}/{\color[rgb]{1,0,0}(\ref{Eq:LSSE_Sol})} to find $\bar{\mathbf{c}}^{\mathcal{A}_n}$ 
        \IF{$\bar{\mathbf{c}}^{\mathcal{A}_n}\geq\mathbf{0}$ holds}
        \STATE Set the values of the elements of $\hat{\bar{\mathbf{c}}}^{\mathtt{CAN}}$, whose indices are in $\mathcal{A}_n$, equal to the values of the corresponding elements in $\bar{\mathbf{c}}^{\mathcal{A}_n}$ and the remaining elements equal to zero;
        \STATE Save $\hat{\bar{\mathbf{c}}}^{\mathtt{CAN}}$ in the candidate set $\mathcal{C}$
        \ELSE
        \STATE Discard $\bar{\mathbf{c}}^{\mathcal{A}_n}$
        \ENDIF
 \ENDFOR
 \STATE Choose {\color[rgb]{0,0,1}$\hat{\bar{\mathbf{c}}}^{\mathtt{ML}}$}/{\color[rgb]{1,0,0}$\hat{\bar{\mathbf{c}}}^{\mathtt{LSSE}}$} equal to that $\hat{\bar{\mathbf{c}}}^{\mathtt{CAN}}$ in the candidate set $\mathcal{C}$ which  {\color[rgb]{0,0,1}maximizes $g(\bar{\mathbf{c}})$}/{\color[rgb]{1,0,0}minimizes $\|\boldsymbol{\epsilon}\|^2$}
\ENDIF
\end{algorithmic}
}
\end{algorithm}

\begin{remk}\label{Remk:ML_Unbiased}
Let us assume a priori that all $L$ taps and the noise mean are  non-zero, i.e., $\bar{\mathbf{c}} > \mathbf{0}$. Thereby, the consistency property of ML estimation \cite[Chapter 4]{BayesianBook} implies that under some regularity
conditions, notably that the likelihood is a continuous function of $\bar{\mathbf{c}}$ and that $\bar{\mathbf{c}}$ is not on the boundary of the parameter set $\bar{\mathbf{c}} \geq \mathbf{0}$, we obtain $\mathbbmss{E}\left\{\hat{\bar{\mathbf{c}}}^{\mathtt{ML}}\right\}\to\bar{\mathbf{c}}$ as $K\to\infty$. In other words, the ML estimator is asymptotically unbiased.  Therefore, for large values of $K$, the ML estimator  becomes  sufficiently  accurate  such  that  none  of  the elements  of  $\hat{\bar{\mathbf{c}}}^{\mathtt{ML}}$ is  zero.  In  this  case,  Algorithm  1  reduces  to
directly solving (\ref{Eq:ML_Sol}) for $\mathcal{A}_n=\mathcal{F}$.
\end{remk}

\subsection{LSSE CIR Estimation}

The LSSE CIR estimator chooses that $\bar{\mathbf{c}}$ which minimizes the sum of the square errors for the observation vector $\mathbf{r}$. Thereby, the error vector is defined as $\boldsymbol{\epsilon} = \mathbf{r} - \mathbbmss{E}\left\{\mathbf{r}\right\} = \mathbf{r} - \mathbf{S} \bar{\mathbf{c}}$ where $\mathbf{S} = [\mathbf{s}_L,\mathbf{s}_{L+1},\dots,\mathbf{s}_K]^T$. In particular, the LSSE CIR estimate can be written as
\begin{IEEEeqnarray}{lll} \label{Eq:LSSE_Estimation}
\hat{\bar{\mathbf{c}}}^{\mathtt{LSSE}} = \underset{\bar{\mathbf{c}}\geq \mathbf{0}}{\mathrm{argmin}} \,\, \| \boldsymbol{\epsilon} \|^2 = \|\mathbf{r} - \mathbf{S} \bar{\mathbf{c}}\|^2.
\end{IEEEeqnarray}
 The square of the norm of the  error vector is obtained as
\begin{IEEEeqnarray}{lll} \label{Eq:LSSE_ErrorNorm}
 \|\boldsymbol{\epsilon}\|^2 & = \mathrm{tr}\left\{ \boldsymbol{\epsilon}\boldsymbol{\epsilon}^T \right\}=  \mathrm{tr}\left\{ (\mathbf{r} - \mathbf{S} \bar{\mathbf{c}}) (\mathbf{r} - \mathbf{S} \bar{\mathbf{c}})^T\right\} \nonumber \\
  &= \mathrm{tr}\left\{ \mathbf{S}^T \mathbf{S} \bar{\mathbf{c}} \bar{\mathbf{c}}^T  \right\} - 2 \mathrm{tr}\left\{ \mathbf{r}^T\mathbf{S}\bar{\mathbf{c}} \right\} + \mathrm{tr}\left\{ \mathbf{r} \mathbf{r}^T \right\},
\end{IEEEeqnarray}
where we used the following properties of the trace:  $\mathrm{tr}\left\{\mathbf{A}\right\}=\mathrm{tr}\left\{\mathbf{A}^T\right\}$ and $\mathrm{tr}\left\{\mathbf{A}\mathbf{B}\right\}=\mathrm{tr}\left\{\mathbf{BA}\right\}$ \cite{TraceDerivative}. The LSSE estimate is given in the following lemma where we use the auxiliary matrix  $\mathbf{S}^{\mathcal{A}_n} = [\mathbf{s}_L^{\mathcal{A}_n} ,\mathbf{s}_{L+1}^{\mathcal{A}_n} ,\dots,\mathbf{s}_K^{\mathcal{A}_n} ]^T$.

\begin{lem}\label{Lem:LSSE}
The LSSE estimator of the CIR for the considered MC channel is given by Algorithm~1 where for a given set $\mathcal{A}_n$,  $\bar{\mathbf{c}}^{\mathcal{A}_n}$ is obtained as  
\begin{IEEEeqnarray}{lll} \label{Eq:LSSE_Sol}
  \bar{\mathbf{c}}^{\mathcal{A}_n} = \left((\mathbf{S}^{\mathcal{A}_n} )^T \mathbf{S}^{\mathcal{A}_n} \right)^{-1}  (\mathbf{S}^{\mathcal{A}_n})^T \mathbf{r}.
\end{IEEEeqnarray}
\end{lem}
\begin{IEEEproof}
The optimization problem in (\ref{Eq:LSSE_Estimation}) is convex  since $\|\boldsymbol{\epsilon}\|^2$ is quadratic in variable $\bar{\mathbf{c}}$, $\mathbf{S}^T \mathbf{S} \succeq 0$ holds, and the feasibility set $\bar{\mathbf{c}}\geq \mathbf{0}$ is linear in $\bar{\mathbf{c}}$ \cite[Chapter 4]{Boyd}. Hence, the constrained convex problem in (\ref{Eq:LSSE_Estimation}) can be solved using a similar methodology as was used to find the ML estimate in Lemma~\ref{Lem:ML}. This leads to Lemma~\ref{Lem:LSSE}.
\end{IEEEproof}

 \begin{remk}
The LSSE estimator employs in fact a linear filter to compute $\bar{\mathbf{c}}^{\mathcal{A}_n}$, i.e., $\bar{\mathbf{c}}^{\mathcal{A}_n} = \mathbf{F}^{\mathcal{A}_n}\mathbf{r}$ where $\mathbf{F}^{\mathcal{A}_n}= \left((\mathbf{S}^{\mathcal{A}_n} )^T \mathbf{S}^{\mathcal{A}_n} \right)^{-1}  (\mathbf{S}^{\mathcal{A}_n})^T$. Moreover, since the training sequence $\mathbf{s}$ is fixed,  matrix  $\mathbf{F}^{\mathcal{A}_n}$ can be calculated offline and then be used  for online CIR estimation. Therefore, the calculation of  $\hat{\bar{\mathbf{c}}}^{\mathcal{A}_n}$ for the LSSE estimator in (\ref{Eq:LSSE_Sol}) is considerably less computationally complex than the computation of $\hat{\bar{\mathbf{c}}}^{\mathcal{A}_n}$ for the ML estimator  in (\ref{Eq:ML_Sol}) which requires solving a system of nonlinear  equations.
\end{remk}

\subsection{CR Lower Bound}

The CR bound is a lower bound on the variance of any unbiased estimator of a deterministic parameter \cite{BayesianBook}. In particular, under some regularity conditions, the covariance matrix of any unbiased estimate of parameter $\bar{\mathbf{c}}$, denoted by $\mathbf{C}(\hat{\bar{\mathbf{c}}})$, satisfies
\begin{IEEEeqnarray}{lll} \label{Eq:CRB_Defin}
\mathbf{C}\left(\hat{\bar{\mathbf{c}}} \right) - \mathbf{I}^{-1} \left(\bar{\mathbf{c}}\right) \succeq 0,
\end{IEEEeqnarray}
where  $\mathbf{I} \left( \bar{\mathbf{c}} \right)$ is the Fisher information matrix of parameter vector $\bar{\mathbf{c}} $ where the elements of $\mathbf{I} \left( \bar{\mathbf{c}} \right)$ are given by
\begin{IEEEeqnarray}{lll} \label{Eq:Fisher_Matrix}
\left[\mathbf{I} \left( \bar{\mathbf{c}} \right)\right]_{i,j} 
&= - \mathbbmss{E}_{\mathbf{r}|\bar{\mathbf{c}}}\left\{  \frac{\partial^2 \mathrm{ln} \{f_{\mathbf{r}}(\mathbf{r}|\bar{\mathbf{c}},\mathbf{s})\} } {\partial\bar{\mathbf{c}}[i] \partial\bar{\mathbf{c}}[j]}  \right\}.
\end{IEEEeqnarray}
We note that for a positive semidefinite matrix, the diagonal elements are non-negative, i.e., $\big[\mathbf{C}(\hat{\bar{\mathbf{c}}}) - \mathbf{I}^{-1} \left(\bar{\mathbf{c}} \right)\big]_{i,i} \geq 0$. Therefore, for an unbiased estimator, i.e., $\mathbbmss{E}\left\{\hat{\bar{\mathbf{c}}}\right\}=\bar{\mathbf{c}}$ holds, with the estimation error vector defined as $\mathbf{e} = \bar{\mathbf{c}} - \hat{\bar{\mathbf{c}}} $,  the CR bound provides the following lower bound on the sum of the expected square errors
\begin{IEEEeqnarray}{ccc} \label{Eq:CRB_CIR}
 \mathbbmss{E}_{\mathbf{r}|\bar{\mathbf{c}}} \left\{ \|\mathbf{e}\|^2\right\}  \geq \mathrm{tr}\left\{ \mathbf{I}^{-1} \left(\bar{\mathbf{c}}\right) \right\} 
 = \mathrm{tr}\left\{ \left[ \sum_{k=L}^{K} \frac{\mathbf{s}_k\mathbf{s}_k^T}{\bar{\mathbf{c}}^T \mathbf{s}_k} \right]^{-1} \right\}. \quad
\end{IEEEeqnarray}

\begin{remk}
We note that the ML and LSSE estimators in Algorithm~1 are biased in general. Hence, the error variances of the ML and LSSE estimates  may fall below the CR bound. However,  as $K\to\infty$, the ML and LSSE estimators become asymptotically unbiased, cf. Remark~\ref{Remk:ML_Unbiased}, and the CR  bound becomes a valid lower bound. The asymptotic unbiasedness   of the proposed estimators is also numerically verified in Section~V, cf. Fig.~\ref{Fig:MeanDet}.
\end{remk}

\section{Training Sequence Design}

In the following, we present two different training sequence designs for CIR estimation in MC systems.

\subsection{LSSE-Based Training Sequence Design}

We first consider a  training sequence design which  minimizes an \textit{upper bound} on the \textit{average} estimation error for the LSSE estimator.  First, we note that for training sequence design, the estimation error has to be averaged over both $\mathbf{r}$ and $\bar{\mathbf{c}}$ since both  are unknown, and hence, have to be modeled as RVs.  Again, we  assume a priori that all $L$  taps and the noise mean are non-zero. Therefore, neglecting the information that  $\bar{\mathbf{c}} \geq \mathbf{0}$ has to hold in (\ref{Eq:LSSE_Estimation}) yields an upper bound on the estimation error for the LSSE estimator. This upper bound is adopted here for the problem of sequence design since the solution of (\ref{Eq:LSSE_Estimation}) after dropping constraint $\bar{\mathbf{c}} \geq \mathbf{0}$ lends itself to an elegant closed-form solution for the estimated CIR given by $\hat{\bar{\mathbf{c}}}^{\mathtt{LSSE}}_{\mathtt{up}}=\left(\mathbf{S}^T \mathbf{S}\right)^{-1}  \mathbf{S} \mathbf{r}$,  which can be used as the basis for either a computer-based search or even a systematic approach to find  good training sequences.  Moreover, this upper bound is tight as $K\to\infty$ since $\hat{\bar{\mathbf{c}}}^{\mathtt{LSSE}}>\mathbf{0}$ holds and we obtain $\hat{\bar{\mathbf{c}}}^{\mathtt{LSSE}}=\hat{\bar{\mathbf{c}}}^{\mathtt{LSSE}}_{\mathtt{up}}$.  In Fig.~\ref{Fig:SeqDes}, we  show numerically that even for short sequence lengths, this upper bound is not loose. 

Defining the estimation error as $\mathbf{e}^{\mathtt{LSSE}}_{\mathtt{up}} = \bar{\mathbf{c}} - \hat{\bar{\mathbf{c}}}^{\mathtt{LSSE}}_{\mathtt{up}}$, the expected square error norm is obtained as
\begin{IEEEeqnarray}{lll} \label{Eq:LSSE_ErrorNorm_Expected1}
 \mathbbmss{E}_{\mathbf{r},\bar{\mathbf{c}}}\left\{\|\mathbf{e}^{\mathtt{LSSE}}_{\mathtt{up}}\|^2\right\} \nonumber \\ = \mathbbmss{E}_{\mathbf{r},\bar{\mathbf{c}}}\left\{\mathrm{tr}\left\{\left(\bar{\mathbf{c}} - \left(\mathbf{S}^T \mathbf{S}\right)^{-1}  \mathbf{S}^T \mathbf{r} \right)\left(\bar{\mathbf{c}} - \left(\mathbf{S}^T \mathbf{S}\right)^{-1}  \mathbf{S}^T \mathbf{r} \right)^T\right\} \right\} \nonumber \\
 =  \mathbbmss{E}_{\mathbf{r},\bar{\mathbf{c}}}\bigg\{ \mathrm{tr}\left\{ \left(\mathbf{S}^T \mathbf{S}\right)^{-1} \mathbf{S}^T \mathbf{r} \mathbf{r}^T \mathbf{S} \left(\mathbf{S}^T \mathbf{S}\right)^{-1} \right\} \nonumber \\
 \qquad - 2 \mathrm{tr}\left\{ \bar{\mathbf{c}}\mathbf{r}^T \mathbf{S} \left(\mathbf{S}^T \mathbf{S}\right)^{-1}  \right\} + \mathrm{tr}\left\{\bar{\mathbf{c}} \bar{\mathbf{c}}^T\right\} \bigg\}.
\end{IEEEeqnarray}
 Next, we calculate the expectation over $(\mathbf{r},\bar{\mathbf{c}})$  in (\ref{Eq:LSSE_ErrorNorm_Expected1})  in two steps, first with respect to $\mathbf{r}$ conditioned on $\bar{\mathbf{c}}$ and then with respect to $\bar{\mathbf{c}}$. To this end, we use  $\mathbbmss{E}_{\mathbf{X}}\left\{\mathrm{tr}\left\{\mathbf{AXB}\right\}\right\} = \mathrm{tr}\left\{\mathbf{A}\mathbbmss{E}_{\mathbf{X}}\left\{\mathbf{X}\right\}\mathbf{B}\right\}$, which is valid for  general matrices $\mathbf{A}$, $\mathbf{B}$, and $\mathbf{X}$,   and  $\mathbbmss{E}_{\mathbf{x}}\left\{\mathbf{x}\mathbf{x}^T\right\}  = \boldsymbol{\lambda}\boldsymbol{\lambda}^T + \mathrm{diag}\{\boldsymbol{\lambda}\}$, which is valid for  multivariate Poisson random vectors $\mathbf{x}$ with covariance matrix $\mathbf{C}(\mathbf{x}) = \mathrm{diag}\{\boldsymbol{\lambda}\}$. Hence, $\mathbbmss{E}\left\{\|\mathbf{e}^{\mathtt{LSSE}}_{\mathtt{up}}\|^2\right\}$ can be calculated as
\begin{IEEEeqnarray}{lll} \label{Eq:LSSE_ErrorNorm_Expected2}
\mathbbmss{E}_{\bar{\mathbf{c}}}\mathbbmss{E}_{\mathbf{r}|\bar{\mathbf{c}}}\left\{\|\mathbf{e}^{\mathtt{LSSE}}_{\mathtt{up}}\|^2\right\}   \nonumber \\
=   \mathbbmss{E}_{\bar{\mathbf{c}}}\bigg\{ \mathrm{tr}\left\{ \left(\mathbf{S}^T \mathbf{S}\right)^{-1} \mathbf{S}^T \left(\mathbf{S}\bar{\mathbf{c}}\bar{\mathbf{c}}^T \mathbf{S}^T\right) \mathbf{S} \left(\mathbf{S}^T \mathbf{S}\right)^{-1} \right\} \nonumber \\
 \qquad - 2 \mathrm{tr}\left\{ \bar{\mathbf{c}}\bar{\mathbf{c}}^T \mathbf{S}^T \mathbf{S} \left(\mathbf{S}^T \mathbf{S}\right)^{-1}  \right\} + \mathrm{tr}\left\{\bar{\mathbf{c}} \bar{\mathbf{c}}^T\right\} \nonumber \\
\qquad + \mathrm{tr}\left\{  \left(\mathbf{S}^T \mathbf{S}\right)^{-1} \mathbf{S}^T \mathrm{diag}\left\{ \mathbf{S}\bar{\mathbf{c}} \right\} \mathbf{S} \left(\mathbf{S}^T \mathbf{S}\right)^{-1} \right\} \bigg\} \qquad \nonumber \\
=\mathbbmss{E}_{\bar{\mathbf{c}}}\left\{ \mathrm{tr}\left\{ \mathbf{S} \left(\mathbf{S}^T \mathbf{S}\right)^{-2} \mathbf{S}^T  \mathrm{diag}\left\{ \mathbf{S}\bar{\mathbf{c}} \right\} \right\} \right\} \qquad \nonumber \\
 = \mathrm{tr}\left\{ \mathbf{S}^T \mathrm{vdiag}\left\{\mathbf{S} \left(\mathbf{S}^T \mathbf{S}\right)^{-2} \mathbf{S}^T\right\}  \boldsymbol{\mu}_{\bar{\mathbf{c}}}^T  \right\},
\end{IEEEeqnarray}
where $\boldsymbol{\mu}_{\bar{\mathbf{c}}} = \mathbbmss{E}_{\bar{\mathbf{c}}}\{\bar{\mathbf{c}}\}$.

\begin{remk}   
The evaluation of the expression in (\ref{Eq:LSSE_ErrorNorm_Expected2}) can be numerically challenging due to the required inversion of matrix $\mathbf{S}^T \mathbf{S}$, especially when one of the eigen-values of $\mathbf{S}^T \mathbf{S}$ is close to zero. One way to cope with this problem is to eliminate all sequences resulting in   close-to-zero eigen-values for matrix  $\mathbf{S}^T \mathbf{S}$ during the search. Formally, we can adopt the following search criterion for training sequence design
\begin{IEEEeqnarray}{lll} \label{Eq:LSSE_Eig}
 \mathbf{s}^* = \underset{\mathbf{s} \in \mathcal{S}}{\mathrm{argmin}} \,\,  \mathrm{tr}\left\{ \mathbf{S}^T \mathrm{vdiag}\left\{\mathbf{S} \left(\mathbf{S}^T  \mathbf{S} \right)^{-2} \mathbf{S}^T\right\}  \boldsymbol{\mu}_{\bar{\mathbf{c}}}^T  \right\}, \quad\,\,
\end{IEEEeqnarray}
where $\mathcal{S}=\left\{\mathbf{s}\big| |x|>  \varepsilon,\,\,\,\,\forall x\in\mathrm{eig}\left\{\mathbf{S}^T\mathbf{S}\right\}  \right\}$ and $\varepsilon$ is a small number which guarantees that the eigen-values of matrix $\mathbf{S}^T  \mathbf{S}$ are not close to zero, e.g., in Section~V, we choose  $\varepsilon=10^{-9}$.
\end{remk}

\subsection{ISI-Free Training Sequence Design}

One simple approach to estimate the CIR is to construct a training sequence such that ISI is avoided during estimation. In this case, in each symbol interval, the receiver will observe molecules  which have been released by the transmitter in only one symbol interval and not in multiple symbol intervals.  To this end, the transmitter releases $N^{\mathtt{Tx}}$  molecules every $L+1$ symbol intervals and remains silent for the rest of the symbol intervals. In particular,  the sequence $\mathbf{s}$ is constructed as follows:
\begin{IEEEeqnarray}{lll} \label{Eq:Seq_ISIfree}
 s[k]=\begin{cases}
 1,\quad &\mathrm{if} \,\,\frac{k-k_0}{L+1}\in \mathbb{Z}\\
 0, & \mathrm{otherwise}
 \end{cases}
\end{IEEEeqnarray}
where $k\in\{1,\dots,K\}$, and $k_0$ is the index of the first symbol interval in which the transmitter releases molecules.   Moreover, for this training sequence, the CIR  can be straightforwardly estimated as
\begin{IEEEeqnarray}{lll} \label{Eq:CIR_ISIfree}
\hat{\bar{c}}^{\mathtt{ISIF}}_l = \frac{1}{|\mathcal{K}_l|} \Big[\sum_{k\in\mathcal{K}_l} \big[ r[k] - \hat{\bar{c}}^{\mathtt{ISIF}}_{\mathtt{n}} \big] \Big]^+, \quad l =1,\dots,L \,\,\, \IEEEyesnumber\IEEEyessubnumber \\
\hat{\bar{c}}^{\mathtt{ISIF}}_{\mathtt{n}} = \frac{1}{|\mathcal{K}_{\mathtt{n}}|} \sum_{k\in\mathcal{K}_{\mathtt{n}}} r[k],     \IEEEyessubnumber
\end{IEEEeqnarray}
where $\mathcal{K}_l=\big\{k|\frac{k-k_0-l+1}{L+1}\in\mathbb{Z}\,\,\wedge\,\,k\in\{1,\dots,K\} \big\}$, $\mathcal{K}_{\mathtt{n}} = \big\{k|\frac{k-k_0-L}{L+1}\in\mathbb{Z} \,\,\wedge\,\,k\in\{1,\dots,K\}  \big\}$, and $[\cdot]^+$ is needed to ensure that all estimated channel taps are non-negative, i.e., $\hat{\bar{\mathbf{c}}}^{\mathtt{ISIF}}\geq\mathbf{0}$ holds.


\section{Performance Evaluation}

In this section, we evaluate the performances of the different estimation techniques and training sequence designs developed in this paper. For simplicity, for the results provided in this section, we generate the CIR $\bar{\mathbf{c}}$ based on (\ref{Eq:Cons_CIR}) and (\ref{Eq:Consentration}). However, we emphasize that the proposed estimation framework is not limited to the particular  channel and receiver models assumed in (\ref{Eq:Cons_CIR}) and (\ref{Eq:Consentration}). We use (\ref{Eq:Cons_CIR}) and (\ref{Eq:Consentration}) only to obtain a  $\bar{\mathbf{c}}$ which is representative of a typical CIR in MC.  In particular, we assume a point source with impulsive molecule release and $N^{\mathtt{Tx}}=10^5$, a fully transparent spherical receiver with radius $45$ nm, and an unbounded environment with $D=4.365\times10^{-10} \,\frac{\text{m}^2}{\text{s}}$ \cite{Arman_AF}. Additionally, we assume that the distance between the transmitter and the receiver is given by $|\mathbf{a}|=|\bar{\mathbf{a}}|+\tilde{a}$ nm  where $|\bar{\mathbf{a}}| = 500$ nm and $\tilde{a}$ is a RV uniformly distributed in the interval $[-\hat{a},\hat{a}]$. The receiver counts the number of molecules once per symbol interval at time $T_{\mathrm{smp}} = {\mathrm{argmax}}_{\,t}\,\,\bar{\mathcal{C}}(\bar{\mathbf{a}},t)$ after the beginning of the symbol interval where $\bar{\mathcal{C}}(\bar{\mathbf{a}},t)$ is computed based on (\ref{Eq:Consentration}). The noise mean is chosen as $\bar{c}_{\mathtt{n}} = 0.5 {\mathrm{max}}_{\,t}\,\,\bar{\mathcal{C}}(\bar{\mathbf{a}},t)$. Furthermore, the symbol duration and the number of taps $L$ are chosen such that $\bar{c}_{L+1}<0.1 \bar{c}_1$.

%

\begin{figure}
  \centering \vspace{-0.6cm}
\resizebox{1\linewidth}{!}{\psfragfig{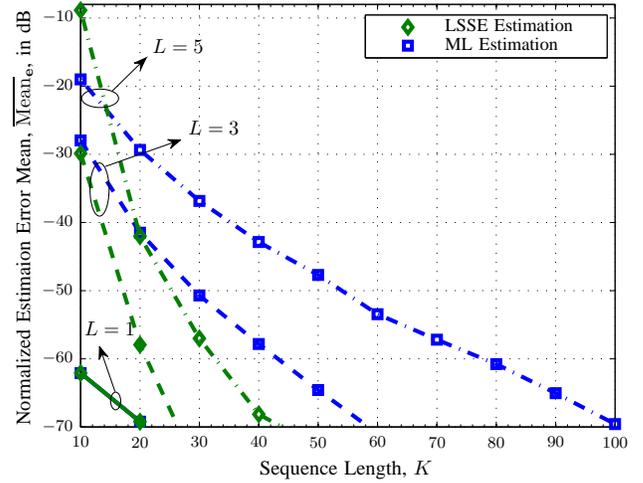}} \vspace{-0.8cm}
\caption{Normalized estimation error mean, $\overline{\mathrm{Mean}}_{\mathbf{e}}$, in dB vs. the training sequence length, $K$, for $L\in\{1,3,5\}$. \vspace{-0.8cm}}
\label{Fig:MeanDet}
\end{figure}

In order to  compare the performances of the considered estimators quantitatively, we define the normalized  mean and variance of the estimation error $\mathbf{e} = \hat{\bar{\mathbf{c}}}-\bar{\mathbf{c}}$ as 
\begin{IEEEeqnarray}{rll} 
 \overline{\mathrm{Mean}}_{\mathbf{e}} \,\,&=  \frac{\left\|\mathbbmss{E}\left\{ \mathbf{e} \right\} \right\|^2}{\|\mathbbmss{E}\left\{\bar{\mathbf{c}}\right\}\|^2} \quad \mathrm{and} \label{Eq:Mean_Norm} \\
 \overline{\mathrm{Var}}_{\mathbf{e}} \,\,&=  \frac{\mathbbmss{E}\left\{\|\mathbf{e}\|^2\right\} - \|\mathbbmss{E}\left\{\mathbf{e}\right\}\|^2 }{\|\mathbbmss{E}\left\{\bar{\mathbf{c}}\right\}\|^2}, \label{Eq:Var_Norm}
\end{IEEEeqnarray}
respectively. In Fig.~\ref{Fig:MeanDet}, we show the normalized mean of the estimation error, $\overline{\mathrm{Mean}}_{\mathbf{e}}$, in dB vs. the training sequence length, $K$, for $L\in\{1,3,5\}$. The training sequences  are constructed by concatenating $n$ copies of the binary sequence $[1 1 0 0 1 0 0 1 0 1]$ of length $10$, i.e., $K=10n$. Furthermore, for clarity of presentation, we assume $\hat{a}=0$ which corresponds to a time-invariant environment with deterministic CIR. The results reported in Fig.~\ref{Fig:MeanDet} are Monte Carlo simulations where each point of the curves is obtained by averaging over $10^6$ random realizations of observation vector $\mathbf{r}$. We observe that the normalized error mean decreases as the sequence length increases. Therefore, the ML and LSSE estimators are biased for short sequence lengths but as the sequence length increases, both the ML and LSSE estimators become asymptotically unbiased, i.e., $\mathbbmss{E}\left\{\hat{\bar{\mathbf{c}}}\right\} \to \bar{\mathbf{c}}$ as $K\to\infty$. Furthermore, from Fig.~\ref{Fig:MeanDet}, we observe that the error mean increases as the number of channel taps increases.

\begin{figure}
  \centering \vspace{-0.4cm}
\resizebox{1\linewidth}{!}{\psfragfig{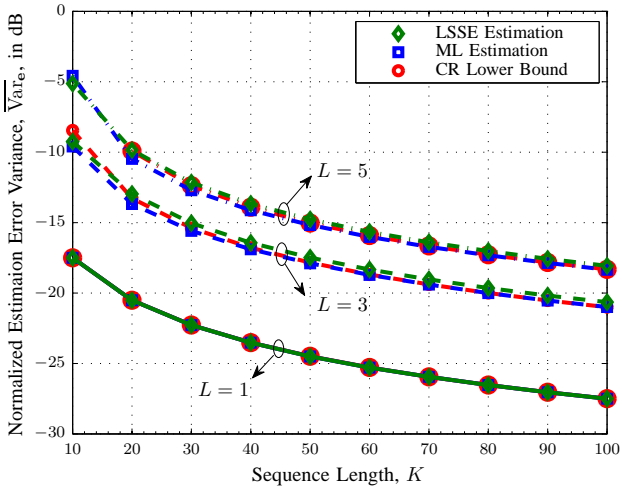}} \vspace{-0.8cm}
\caption{Normalized estimation error variance, $\overline{\mathrm{Var}}_{\mathbf{e}}$, in dB vs. the training sequence length, $K$, for $L\in\{1,3,5\}$. \vspace{-0.3cm}}
\label{Fig:DetermCIR}
\end{figure}

In Fig.~\ref{Fig:DetermCIR}, we show the normalized estimation error variance, $\overline{\mathrm{Var}}_{\mathbf{e}}$, in dB vs. the training sequence length, $K$, for $L\in\{1,3,5\}$.  The parameters used in Fig.~\ref{Fig:DetermCIR} are identical to those used in Fig.~\ref{Fig:MeanDet}.  As expected, the variance of the estimation error decreases with increasing  training sequence length. Moreover, for $L\in\{3,5\}$, we observe that the variance of the estimation error for the LSSE estimator is slightly higher than that for the ML estimator,  whereas for $L=1$, the variance of the estimation error  for the LSSE estimator coincides with that of the ML estimator. These results suggest  that the simple LSSE estimator provides a favorable complexity-performance tradeoff for CIR estimation in the considered MC system. For short sequence lengths, the variances of the ML and LSSE estimators can even be  lower than the CR bound as these estimators are biased and the CR bound is a valid lower bound only for unbiased estimators, see Fig.~\ref{Fig:MeanDet}. However, as $K$ increases, both the ML and LSSE estimators become asymptotically  unbiased, see Fig.~\ref{Fig:MeanDet}. Fig.~\ref{Fig:DetermCIR} shows that, for large $K$, the error variance of the ML estimator coincides with the CR bound and the error variance of the LSSE estimator is very close to the CR bound.  We note that for the adopted training sequence,  the matrix inversion required in (\ref{Eq:CRB_CIR}) cannot be computed for $K=10$ and $L=5$ since matrix $\sum_{k=L}^{K} \frac{\mathbf{s}_k\mathbf{s}_k^T}{\bar{\mathbf{c}}^T \mathbf{s}_k} $ has one zero eigen-value. Therefore, we do not report the value of the CR bound for this case in Fig.~\ref{Fig:DetermCIR}.

Next, we investigate the performances of the optimal and ISI-free training sequence designs developed in Section~IV. Here, we employ a computer-based search to find the optimal sequence  based on the criterion in (\ref{Eq:LSSE_Eig}) where $\varepsilon=10^{-9}$. We consider short sequence lengths, i.e., $K\leq 20$, due to exponential increase of the computational complexity of the exhaustive search with respect to the sequence length. Moreover, since there are $L+1$ unknown parameters, we require at least $L+1$ observations for estimation, i.e., $K-L+1\geq L+1$ or equivalently $K\geq 2L$. In Table~I, we present the optimal sequences obtained for  $L\in\{1,2,3,4,5\}$, $K\in\{10,16\}$, and $\hat{a}=100$ nm.  We note that the optimal sequence which is obtained from  (\ref{Eq:LSSE_Eig}) is not unique and only one of the optimal sequences is shown in Table~I for each value of $K$ and $L$.  The optimal sequences shown in blue font in Table~I are identical to the ISI-free sequences proposed in (\ref{Eq:Seq_ISIfree}). In particular, for $L=1$, the optimal sequences for both $K=10$ and $16$ are ISI-free,  whereas for $L>1$,  none of the optimal sequences is ISI-free. 

\begin{figure}
  \centering\vspace{-0.5cm}
\resizebox{1\linewidth}{!}{\psfragfig{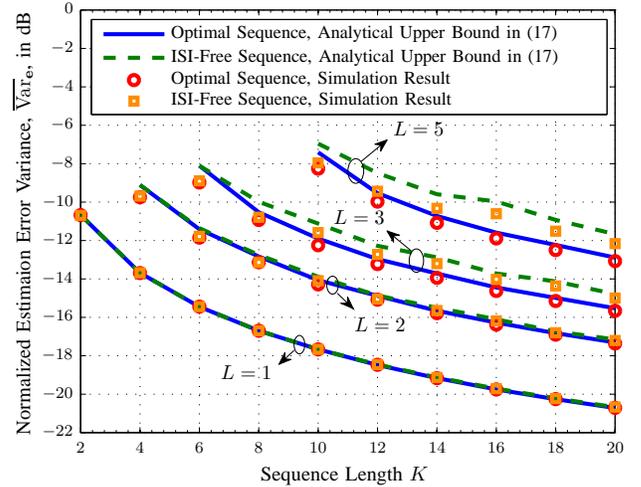}} \vspace{-0.9cm}
\caption{Normalized LSSE estimation error variance, $\overline{\mathrm{Var}}_{\mathbf{e}}$, in dB vs. the training sequence length, $K$, for $L\in\{1,2,3,5\}$. \vspace{-0.6cm}}
\label{Fig:SeqDes}
\end{figure}

In Fig.~\ref{Fig:SeqDes}, we show the normalized LSSE estimation error, $\overline{\mathrm{Var}}_{\mathbf{e}}$, in dB vs. the training sequence length, $K$, for $L\in\{1,2,3,5\}$ and $\hat{a}=100$ nm. Thereby, we report the analytical results for the upper bound in (\ref{Eq:LSSE_ErrorNorm_Expected2}) and  Monte Carlo simulation results for $10^6$ random realizations. Fig.~\ref{Fig:SeqDes} confirms that  (\ref{Eq:LSSE_ErrorNorm_Expected2}) is a tight upper bound even for short sequence lengths. Moreover, we observe from Fig.~\ref{Fig:SeqDes} that the performance of the  ISI-free sequence coincides with that of the optimal sequence for all sequence lengths when $L=1$, and for $L>1$, the difference between the error variances of the ISI-free sequence and the optimal sequence increases as $L$ increases. This result suggests that for MC channels with small numbers of taps, a simple ISI-free training sequence is a suitable option.  Furthermore, as expected, the estimation error decreases with increasing  training sequence length.
 
\begin{table}
\label{Table:OptSeq}
\caption{Examples of Optimal LSSE Sequences  Obtained by a Computer-Based Search for  $L\in\{1,\dots,5\}$ and $K\in\{10,16\}$.} 
\begin{center}
\scalebox{0.88}{
\begin{tabular}{|| c | c | c  ||}
  \hline                  
      &  $K=10$ &  $K=16$   \\ \hline
$L=1$  &   {\color[rgb]{0,0,1}$\mathbf{s}^* = [ 1     0     1     0     1     0     1     0     1     0]^T$} &  
{\color[rgb]{0,0,1} $\,\,\mathbf{s}^* = [ 0     1     0     1     0     1     0     1     0     1     0     1     0     1     0     1 ]^T$ }\\ \hline
$L=2$  &   $\mathbf{s}^* = [0     0     1     0     0     0     1     1     1     0]^T$ & 
  $\mathbf{s}^* = [0     1     1     0     0     1     1     1     0     1     0     0     0     0     0     1]^T$ \\ \hline
 $L=3$  &   $\mathbf{s}^* = [0     1     0     0     0     0     1     1     0     1 ]^T$ & 
 $\mathbf{s}^* = [0     1     0     1     1     0     1     1     0     1     1     0     0     0     0     0]^T$ \\ \hline
  $L=4$  &   $\mathbf{s}^* = [1     0     1     0     1     1     0     0     0     0]^T$ & 
 $\mathbf{s}^* = [1     1     1     1     1     0     0     0     0     1     0     0     0     1     0     0]^T$ \\ \hline 
  $L=5$  &   $\mathbf{s}^* = [0     1     1     0     1     0     0     0     1     0]^T$ & 
 $\mathbf{s}^* = [1     0     0     1     0     1     0     0     1     1     0     0     0     0     0     0]^T$ \\ \hline 
\end{tabular}
}
\end{center}
\vspace{-0.5cm}
\end{table}

\section{Conclusions}

In this paper, we developed a training-based CIR estimation framework which enables the acquisition of the CIR based on the observed number of molecules at the receiver due to emission of a  sequence of known numbers of molecules by the transmitter.  We derived the optimal ML estimator, the suboptimal LSSE  estimator, and the CR lower bound. Furthermore, we studied both an optimal and a suboptimal training sequence design for the considered MC system.  Simulation results confirmed the analysis and compared the performance of the proposed estimation techniques with the CR lower bound. 

\bibliographystyle{IEEEtran}
\bibliography{Ref_09_10_2015}

\end{document}